\documentclass[10pt,a4paper,oneside]{article}
\pdfoutput=1
\usepackage[english]{babel}
\usepackage{amsmath,amsfonts,amssymb,bm}\interdisplaylinepenalty=2500

\usepackage{graphicx,color,rotating,epstopdf}
\graphicspath{{figs/}}					
\def\LatexGraphicsPath{./figs/}			
\def\PrintGraphicFileName{1}			
\newcommand{\namedgraphics}[2]{
	\parbox{\textwidth}{%
	\ifnum\PrintGraphicFileName>0\rotatebox{90}{\smash{\ttfamily\scriptsize\raisebox{0.8em}{#2}}}\fi%
	\hspace*{\fill}\includegraphics[scale=#1]{#2}\hspace*{\fill}}}
\newcommand{\namedgraphicsOneCol}[2]{
	\parbox{\columnwidth}{%
	\ifnum\PrintGraphicFileName>0\rotatebox{90}{\smash{\ttfamily\scriptsize\raisebox{0.4em}{#2}}}\fi%
	\hspace*{\fill}\includegraphics[scale=#1]{#2}\hspace*{\fill}}}
\newcommand{\namedcombographics}[2]{
	\parbox{\textwidth}{%
	 \ifnum\PrintGraphicFileName>0\rotatebox{90}{\smash{\ttfamily\scriptsize\raisebox{0.8em}{#2}}}\fi%
	\hspace*{\fill}\scalebox{#1}{
	\ifnum\pdfoutput=0\input{\LatexGraphicsPath#2.pstex_t}
	\else\input{\LatexGraphicsPath#2.pdftex_t}\fi}\hspace*{\fill}}}
\newcommand{\namedlatexgraphics}[2]{
	\parbox{\textwidth}{%
	 \ifnum\PrintGraphicFileName>0\rotatebox{90}{\smash{\ttfamily\scriptsize\raisebox{0.8em}{#2}}}\fi%
	\hspace*{\fill}\scalebox{#1}{
	\input{\LatexGraphicsPath#2.latex}}\hspace*{\fill}}}

\title{Determination of Phase Noise Spectra in\\Optoelectronic Microwave Oscillators:\\a Langevin Approach}
\author{Y. Kouomou Chembo\thanks{FEMTO-ST, Dept.\ of Optics.  Corresponding author, e-mail: yanne.chembo@femto-st.fr and ckyanne@yahoo.fr.}
\thanks{Y.K.C. acknowledges a research fellowship from the R\'egion de Franche-Comt\'e in France, and a financial support from MEC (Spain) and FEDER under projects TEC2006-10009 (PhoDeCC) and FIS2007-60327 (FISICOS).},
K. Volyanskiy\thanks{FEMTO-ST, Dept.\ of Optics, and St. Petersburg State University of Aerospace Instrumentation.},\\
L. Larger\thanks{FEMTO-ST, Dept.\ of Optics.},
E. Rubiola\thanks{FEMTO-ST, Dept.\ of Time and Frequency.  Home page http://rubiola.org.}~ and 
P. Colet\thanks{Instituto de F\'isica Interdisciplinar y Sistemas Complejos IFISC (CSIC-UIB), Palma de Mallorca, Spain.}
\\[4em]\includegraphics[width=0.35\textwidth]{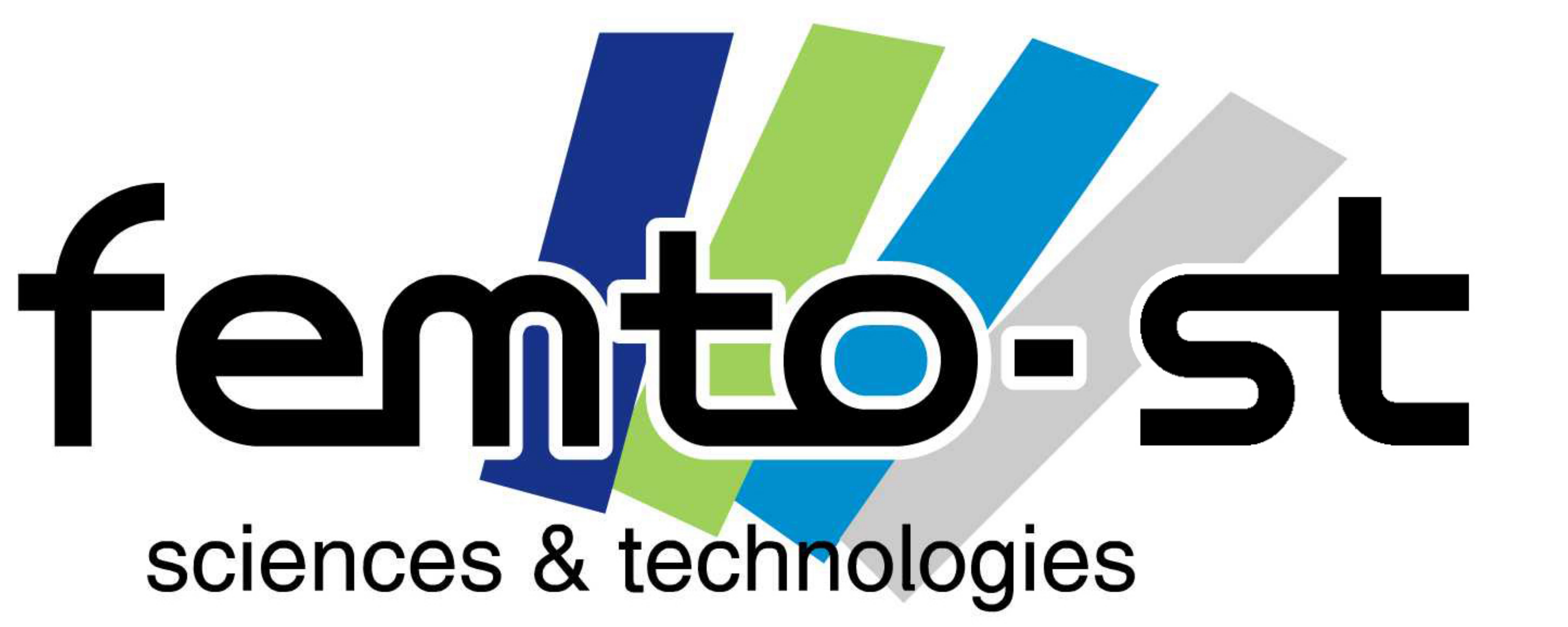}\\[0.5em]
\small FEMTO-ST Institute\\[-0.5ex]
\small CNRS and Universit\'e de Franche Comt\'e, 
\small Besan\c{c}on, France\\[1.5em]}

\date{\small\today}
\def\myheaders{Y. K. Chembo \& al. \hfill\today\quad}
\begin{document}
\maketitle

\begin{abstract}
We introduce a stochastic model for the determination of phase noise in optoelectronic oscillators.
After a short overview of the main results for the phase diffusion approach in autonomous oscillators, an extension is proposed for the case of optoelectronic oscillators where the microwave is a limit-cycle originated from a bifurcation induced by nonlinearity and time-delay.
This Langevin approach based on stochastic calculus is also successfully confronted with experimental measurements.
\end{abstract}


\clearpage
\tableofcontents
\clearpage

\section{Introduction}
Optoelectronic oscillators (OEOs) combine a nonlinear modulation of laser light with optical storage to generate ultra-pure microwaves for lightwave telecommunication and radar applications \cite{YaoMalekiJOSAB,YaoMalekiJQE}. Their principal specificity is their extremely low phase noise, which can be as low as $-160$~dBrad$^2$/Hz at $10$~kHz from a $10$~GHz carrier.
Despite some interesting preliminary investigations, the theoretical determination of phase noise in OEOs is still a partially unsolved problem. The qualitative features of this phase noise spectrum can be recovered using some heuristical guidelines or rough approximations, but however, a rigorous theoretical background is still lacking.

There are several reasons which can explain that absence of theoretical background.
A first reason is that before refs. \cite{OL}, there was no time domain model to describe such systems, so that stochastic analysis could not be used to perform the phase noise study. Moreover, unlike most of oscillators, the OEO is a delay-line oscillator, and very few had been done to study the effect phase noise on time-delay induced limit-cycles. Finally, the OEO is subjected to multiple noise sources, which are sometimes non-white, like the flicker (also referred to as ``$1/f$") noise which is predominant around the microwave carrier.

The objective of this work is to propose a theoretical study where all these features are taken into account. The plan of the article is the following.
In Section \ref{phasediffusion}, we present the phase diffusion approach in autonomous systems. It is a brief review where the fundamental concepts of phase diffusion are recalled, and where some important earlier contributions are highlighted.
Then, we derive in Section \ref{phaseSDDE} a stochastic delay-differential equation for the phase noise study.
We show that for our purpose, the global interaction of noise with the system can be decomposed into two contributions, namely an additive and a multiplicative noise contribution.
Section \ref{belowth} is devoted to the study of the noise spectrum below threshold. It will appear that the spectrum below threshold will not only be important to validate the stochastic model, but also that it enables an accurate calibration of additive noise.
In Section \ref{aboveth}, we address the problem of phase noise when there is a microwave output using Fourier analysis, and we show that it is possible to have an accurate image of the phase noise spectrum in all frequency ranges. The last section concludes the article.

\section{The phase diffusion approach in autonomous oscillators}
\label{phasediffusion}

\subsection{Fundamental concepts}

For an ideal (noise-free) oscillator, the Fourier spectrum is a collection of Dirac peaks, standing for the fundamental frequency and its harmonics.
The effect of amplitude white noise is to add a flat background, while the peaks do keep their zero linewidth; it is the effect of phase noise to widen the linewidth of these peaks.

Some pioneering papers on the topic of phase noise in autonomous oscillators using stochastic calculus had been published forty years ago \cite{LaxV}.
In particular, it was demonstrated that a general framework to study the problem of phase noise in a self-sustained oscillator could be built using some minimalist assumptions. The first point is that a strong nonlinearity is an essential necessity in oscillators, in the sense that nonlinearity can not be regarded as small because it controls the operating level of the oscillator. The second important point is that the phase is only {\em neutrally} stable, so that quasilinear methods which assume that fluctuations from some operating point are small (linearization techniques) can not be applied {\em directly}.

The phase is neutrally stable as a consequence of the phase-invariance of autonomous
oscillators. In other words, limit-cycles are stable against amplitude perturbations, while there is no mechanism able to stabilize the phase to a given value: hence, phase perturbations are undamped, but they do not diverge exponentially, though. In a noise free oscillator, the ``stroboscopic" state point on the limit-cycle is immobile, but in the presence of noise, it moves randomly along the limit-cycle: in other words, the phase of the oscillator undergoes a {\em diffusion} process, in all points similar to a one-dimensional Brownian motion. In the most simple case, the random fluctuations of the phase are referred to as a {\em Wiener process}, obeying an equation of the kind $\dot{\varphi} =  \xi (t)$, where $\xi$ is a Gaussian white noise with autocorrelation $\left< \xi(t) \xi(t')\right> = 2D \delta(t-t')$, while $D$ is a parameter referred to as the {\em diffusion constant}. It can be demonstrated that the phase variance diverges linearly as $\left< \varphi^2(t)\right>=2Dt$, and the single-side band phase noise spectrum (in dBc/Hz) explicitly reads ${\cal L}(\omega)=2D/[D^2+\omega^2]$, so that $D$ is the unique parameter characterizing all the statistical and spectral features of phase fluctuations.

\begin{figure}
\begin{center}
\includegraphics[width=6cm]{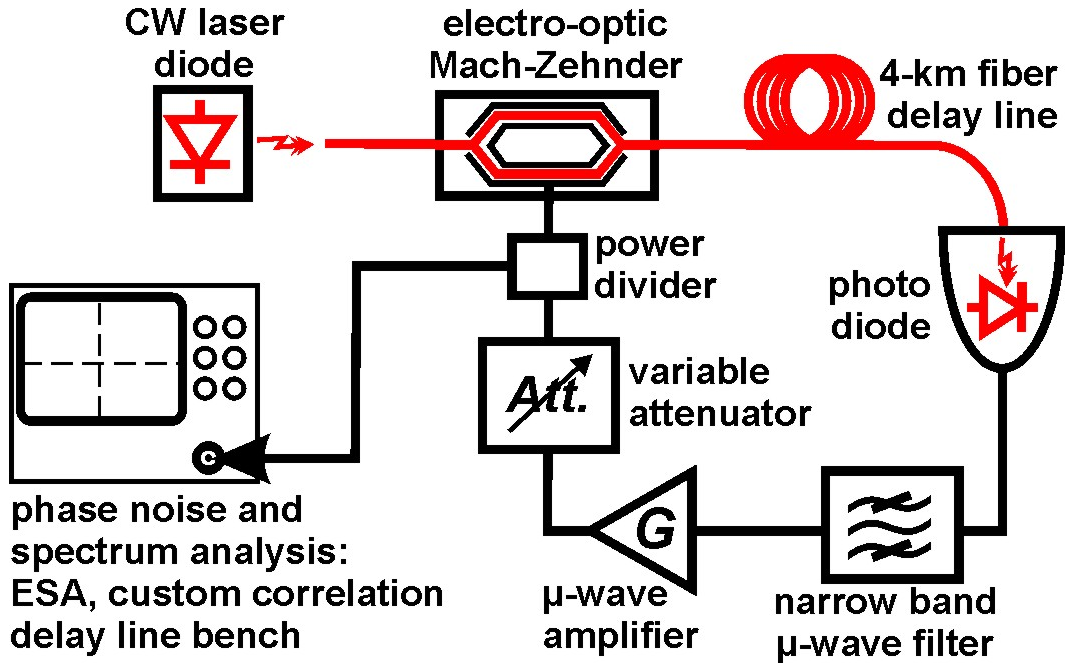}
\end{center}
\caption[Setup]
{ \label{Setup} Experimental set-up.}
\end{figure}

\subsection{The unifying theory of Demir, Mehrota and Roychowdhury}

On the base of earlier works by Lax \cite{LaxV} and K\"artner \cite{Kartner}, Demir, Mehrota and Roychowdhury have proposed few years ago a unifying theory of phase noise in self-sustained oscillators subjected to white noise sources \cite{DemirMC}. Their approach, which had later been extended by Demir to the case colored noise sources \cite{Demir}, relies on stochastic calculus. The principal point of their contribution was the introduction of a decomposition of phase and amplitude noise through a projection onto the periodic time-varying eigenvectors
(the so called {\em Floquet eigenvectors}; also see ref. \cite{Coram}), and they proved that it provides the correct solution to the problem.

Demir {\em et al} have shown that if the sources of noise are Gaussian and white, the phase noise around the fundamental peak (and its harmonics) has a Lorentzian lineshape, and therefore is fully determined by an ``effective" diffusion constant $D_{eff}$ which is the unique parameter needed for the phase noise determination. However, if the Demir {\em et al} theory has the great and essential advantage of mathematical rigorousness, its principal drawback is that exactitude is obtained at the expense of simplicity: the calculation of $D_{eff}$ is very complex, as it requires an accurate determination of all the time-varying eigenvectors related to the autonomous flow. In general this task can only be performed numerically using quite complicated algorithms, and this lack of flexibility explains why this method is scarcely used in the phase noise studies available in the literature.
The key challenge for the study of phase noise in OEO would be provide an accurate description of the phase noise spectrum, while avoiding the determination of Floquet eigenvectors, which is an extremely complicated task in delayed system.

\section{Application of the phase diffusion approach to OEOs: stochastic delay-differential equations}
\label{phaseSDDE}

The OEO under study is organized in a single-loop architecture as depicted in Fig. \ref{Setup}.
The oscillation loop consists of:
$(i)$   A wideband integrated optics LiNbO$_3$ Mach-Zehnder (MZ) modulator,
        seeded by a continuous-wave semiconductor laser of optical power $P$;
        the modulator is characterized by a half-wave voltage $V_\pi = 4$~V.
$(ii)$  A thermalized $4$~km fiber performing a time delay of $T=20$ $\mu$
        on the microwave signal carried by the optical beam; the corresponding free spectral range is $\Omega_T/2\pi= 1/T= 50$~kHz.
$(iii)$ A fast photodiode with a conversion factor $S$.
$(iv)$  A narrow band microwave radio-frequency (RF) filter, of central frequency $F_0=\Omega_0/2\pi=10$~GHz, and $-3$~dB bandwidth of
	    $\Delta F=\Delta\Omega/2\pi=50$~MHz;
$(v)$   A microwave amplifier with gain $G$.
$(vi)$  A variable attenuator, in order to scan the gain.
$(vii)$  All optical and electrical losses are gathered in a single attenuation factor $\kappa$.

The dynamics of the microwave oscillation can therefore be described in terms of
the dimensionless variable $x(t)=\pi V(t)/2V_\pi$  whose dynamics obeys \cite{OL}
\begin{eqnarray}
x+ \tau \frac{dx}{dt}+\frac{1}{\theta} \int_{t_0}^{t} x(s) ds = \beta \cos^{2} [x(t-T)+\phi] \, ,
\label{OEO_original}
\end{eqnarray}
where $\beta=\pi\kappa S GP/2V_{\pi}$ is the normalized loop gain,
$\phi=\pi V_B/2V_{\pi}$ is the Mach-Zehnder offset phase, while
$\tau=1/\Delta\Omega$ and $\theta=\Delta\Omega/\Omega_0^2$
are the characteristic timescale parameters of the bandpass filter. Since we are interested by single-mode microwave oscillations, the solution of Eq. (\ref{OEO_original}) can be expressed under the form
\begin{eqnarray}	
x(t)= \frac{1}{2} {\cal A}(t)\, e^{i \Omega_0 t}  + \frac{1}{2} {\cal A}^*(t)\, e^{-i \Omega_0 t} \, ,
\label{expr_x_cal_A}
\end{eqnarray}	
where ${\cal A}(t)=A(t) \, \exp[i \psi(t)]$ is the slowly varying amplitude of the microwave $x(t)$.
We can significantly simplify the right-hand side term of Eq.~(\ref{OEO_original}) because
the cosine of a sinusoidal function of frequency $\Omega_0$ can be Fourier-expanded in harmonics
of $\Omega_0$. In other words, since $x(t)$ is nearly sinusoidal around
$\Omega_0$, then the Fourier spectrum of $\cos^2 [x(t-T)+\phi]$ will be sharply distributed around the harmonics of
$\Omega_0$ using the relationship $\cos^2 z = [1+ \cos 2z]/2$ and the Jacobi-Anger expansion
\begin{eqnarray}
e^{iz \cos \alpha } = \sum_{n=-\infty}^{+\infty } i^n {\rm J_n}(z)e^{i n \alpha } \, ,
\label{jacobi_anger}
\end{eqnarray}
where ${\rm J_n}$ is the $n$-th order Bessel function of the first kind.
Hence, since the filter of the feedback loop is narrowly resonant around $\Omega_0$, it can be demonstrated that discarding all the spectral components of the signal except the fundamental is an excellent approximation, so that Eq.~(\ref{OEO_original}) can be rewritten as
\begin{eqnarray}
&& x+ \frac{1}{\Delta \Omega} \frac{dx}{dt}+\frac{\Omega_0^2}{\Delta \Omega} \int_{t_0}^{t} x(s) ds = -\beta \sin 2 \phi  \nonumber \\
&& \times {\rm J_1} [2 |{\cal A}(t-T)|] \, \cos[ \Omega_0 (t-T)+ \psi(t-T)] \, .
\label{OEO_J1}
\end{eqnarray}
In order to include noise effects in this equation, we will consider two main noise contributions in this system.

The first contribution is an {\em additive noise}, corresponding to random environmental and internal fluctuations which are uncorrelated from the eventual existence of a microwave signal. The effect of this noise can be accounted for by addition as a Langevin forcing term, to be added in the right-hand side of Eq.~(\ref{OEO_J1}).
This additive noise can be assumed to be spectrally white, and since we are interested by its intensity around the carrier frequency $\Omega_0$, it can be explicitly written as
\begin{eqnarray}	
\xi_a(t) =  \frac{1}{2} \zeta_a(t) e^{i \Omega_0 t} +\frac{1}{2} \zeta_a^*(t) e^{-i \Omega_0 t}  \, ,
\label{decomp_noises}
\end{eqnarray}	
where $\zeta_a(t)$ is a complex Gaussian white noise, whose correlation is
$\left< \zeta_{a}(t)\zeta_{a}^*(t')\right> = 4 D_a \delta(t-t')$, so that the corresponding power density spectrum is
$|\tilde{\xi}_a(\omega)|^2 = 2 D_a$.

The second contribution is a {\em multiplicative noise} due to a noisy loop gain. Effectively, the normalized gain parameter
explicitly reads
\begin{eqnarray}
\gamma =   \beta \sin 2 \phi =  \frac{\pi}{2} \frac{\kappa S G P}{V_{\pi_{RF}}} \, \sin\left[ \pi \, \frac{V_B}{V_{\pi_{DC}}}\right]   \, .
\label{gamma_explicit}
\end{eqnarray}
If all the parameters of the system are noisy (i.e., we replace $\kappa$ by $\kappa+\delta \kappa(t)$, $S$ by $S+\delta S(t)$, etc.),
then the gain $\gamma$ may be replaced in Eq.~(\ref{OEO_J1}) by $\gamma + \delta \gamma (t) $, where the $\delta \gamma (t)$ is the overall gain fluctuation. We therefore introduce the dimensionless multiplicative noise
\begin{eqnarray}
\eta_m (t) = \frac{\delta \gamma (t)}{\gamma} \, ,
\label{multiplicative_noise}
\end{eqnarray}
which is in fact the relative gain fluctuation. In the OEO configuration, we have $\eta_m (t) \ll 1$. This noise is in general spectrally complex, as it is the sum of noise contributions which are very different (noise from the photodetector, from the amplifier, etc.). In agreement with the usual noise spectrum of amplifiers and photodetectors, we will here consider that this multiplicative noise is flicker (i.e., varies as $1/f$) near the carrier, and white above a certain knee-value. We therefore assume the following empirical noise power density
\begin{eqnarray}
|\tilde{\eta}_m(\omega)|^2 = 2D_m \, \left[ 1+ \frac{\Omega_{H}}{\omega + \Omega_{L}}   \right] \, ,
\label{mult_noi_freq_dep}
\end{eqnarray}
where $\Omega_{L}$ is the low corner frequency of the flicker noise, while $\Omega_{H}$ is the high corner frequency.
More precisely, we consider that the noise is white below $\Omega_L$ and above $\Omega_H$, while it remains flicker in between. Typically, we may consider $\Omega_{L}/2\pi < 1$ Hz  and  $\Omega_{H}/2\pi > 10$ kHz, so that the flicker noise is extended over a frequency span of more than 4 orders of magnitude. \\

To avoid the integral term of Eq. (\ref{OEO_J1}) which is complicated to manage analytically, it is mathematically
convenient to use the intermediate integral variable
\begin{eqnarray}	
u(t) = \int_{t_0}^{t}x(s) \, ds
     = \frac{1}{2} {\cal B}(t)\, e^{i \Omega_0 t}  + \frac{1}{2} {\cal B}^*(t)\, e^{-i \Omega_0 t}  \, ,
\label{expr_u_integr}
\end{eqnarray}	
which is also nearly sinusoidal with a zero mean value.
Using Eqs. (\ref{OEO_J1}), (\ref{decomp_noises}) and (\ref{multiplicative_noise}), it can be shown that the slowly-varying amplitude ${\cal B}(t)$  obeys the stochastic equation
\begin{eqnarray}
&& \{ \ddot{{\cal B}} + (\Delta \Omega  + 2 i \Omega_0) \dot{{\cal B}} + i \Omega_0 \, \Delta \Omega \, {\cal B} \}e^{i \Omega_0 t}
   + {\rm c.c.} \nonumber \\
&& = -2 \Delta \Omega  \gamma \left[ 1+ \eta_m(t) \right]\,
\, \left[ \frac{1}{2}e^{i \Omega_0 (t-T)}e^{i\psi_T}  + {\rm c.c.} \right] \nonumber \\
&& \times {\rm J_1}[2 |\dot{{\cal B}}_T+i \Omega_0 {\cal B}_T|]+
2 \Delta \Omega  \left[ \frac{1}{2}\zeta_a(t)e^{i \Omega_0 t} + {\rm c.c.} \right] \, , \nonumber \\
\label{eq_calB_stoch_1}
\end{eqnarray}
where {\rm c.c.} stands for the complex conjugate of the preceding term.
We can assume $ |\ddot{{\cal B}}| \ll \Delta \Omega  |\dot{{\cal B}}|$ and $ |\dot{{\cal B}}| \ll \Omega_0  |{{\cal B}}|$; the relationship $x(t) = \dot{u}(t)$ therefore gives ${{\cal A}} \simeq i\Omega_0 {{\cal B}}$, so that we can finally derive from Eq.~(\ref{eq_calB_stoch_1}) the following stochastic equation for the slowly varying envelope ${\cal A}(t)$
\begin{eqnarray}
\dot{{\cal A}}&=& - \mu  e^{i\vartheta } {\cal A}
+ 2 \gamma \mu e^{i\vartheta }   \left[ 1+ \eta_m(t) \right] \, {\rm Jc_1}[2 |{\cal A}_T|]  {\cal A}_T \nonumber \\
&& + \mu e^{i\vartheta } \zeta_a(t)\, ,
\label{eqt_A_stoch}
\end{eqnarray}
where ${\rm Jc}_1 (x)= {\rm J}_1(x)/x$ is the first order \emph{Bessel cardinal} function of the first kind. The phase condition has been set to $e^{-i \Omega_0 T}=-1$, so that the dynamics of interest is restricted to the case $\gamma \geq 0$.
The key parameters of this equation are
\begin{eqnarray}
\mu = \frac{\Delta \Omega /2}{\sqrt{1+(1/2Q)^{2}}} \,\,\,{\rm and}\,\,\, \vartheta = \arctan \left[ \frac{1}{2Q}\right]\, ,
\label{def_param_SDDE}
\end{eqnarray}
where $Q=\Omega_0/\Delta \Omega = 200$ is the quality factor of the RF filter. Since $Q \gg 1$, we may simply consider that $\mu \simeq \Delta \Omega/2$ and $ \vartheta \simeq 1/2Q$.  The complex term $\mu e^{i\vartheta }$ is a kind of ``filter operator'', which can be simply equated to the half-bandwidth $\Delta \Omega /2$ when the $Q$-factor of the filter is sufficiently high, as it was done in ref. \cite{OL}. It is also noteworthy that in the complex amplitude equation (\ref{eqt_A_stoch}), the initial multiplicative noise remains a {\em real} variable, while the additive noise becomes {\em complex}.

We had recently shown, in agreement with the experiment, that the OEO has three fundamental regimes \cite{OL}. For $\gamma  <1$, the system does not oscillate and the trivial fixed point is stable; for $1 \leq \gamma < 2.31$, the system sustains a pure microwave oscillation, with a constant amplitude and frequency; and at last, for $\gamma \geq 2.31$, the system enters into a regime where the amplitude of the microwave is unstable, and turns to be nonlinearly modulated. We can consider that this phenomenology is still correct as long as $Q \gg 1$.
With the aid of the stochastic delay-differential equation ruling the dynamics of ${\cal A}$, we may now derive analytically the power spectrum density of the oscillator, below and above threshold. However, it should be stressed that in {\em all} cases, stochastic variables should be manipulated with respect to the rules of stochastic calculus when an integral/differential transformation is applied to them.

\section{Noise power density spectrum below threshold (\boldmath$\gamma < 1$)}
\label{belowth}

In general, no interest is paid to the study of the noise power density spectrum below threshold in OEOs. This lack of interest can be explained by the fact that there is no oscillation in this regime, and the system randomly fluctuates around the trivial equilibrium.
However, as we will further see, this regime is particularly interesting because it enables to understand how the noise interacts with the system.

From the stability theory of delay-differential equations with complex coefficients,
the deterministic solution of Eq.~(\ref{eqt_A_stoch}) below threshold is the trivial fixed point ${\cal A}=0$.
After linearization around this solution, Eq.~(\ref{eqt_A_stoch}) can simply be rewritten as
\begin{eqnarray}
\dot{{\cal A}}= - \mu  e^{i\vartheta } {\cal A} + \gamma  \mu e^{i\vartheta }  {\cal A}_T + \mu e^{i\vartheta } \zeta_a(t) \, ,
\label{A_Bel_Th}
\end{eqnarray}
where we have used ${\rm Jc_1}(0)=1/2$.
This equation indicates that the multiplicative noise has no significative influence below threshold, because the product
$\eta_m {\cal A}_T$ is a second-order term. Therefore, the noise power below threshold is {\em essentially} determined by additive noise.

Equation~(\ref{A_Bel_Th}) is linear with constant coefficients: hence, the power density spectrum can directly be obtained as
\begin{eqnarray}
 |\tilde{{\cal A}}(\omega)|^2
=  \frac{4 \mu^2 D_a}{\left| i \omega e^{-i\vartheta} + \mu (1-\gamma e^{- i \omega T}) \right|^2}   \, .
\label{A_Bel_Th_Four_2}
\end{eqnarray}

One can determine the total output power below threshold due to the white noise fluctuations in the system through the formula
\begin{eqnarray}
P_{\gamma}= \left(\frac{2 V_{\pi}}{\pi}\right)^2 \frac{\left< |{\cal A}(t)|^2\right>}{2R}  \, ,
\label{noise_pow_true}
\end{eqnarray}
where $R$ is the output impedance (in our case, $R=50$~$\Omega$). The dimensionless power $\left< |{\cal A}(t)|^2\right>$ can not be calculated analytically for $\gamma \neq 0$: it can nevertheless be determined either by numerical simulation of Eq.~(\ref{A_Bel_Th}), or through a numerical computation of the integral $\frac{1}{2\pi} \int_{-\infty}^{+\infty}|\tilde{{\cal A}}(\omega)|^2\, d\omega$, where $\tilde{{\cal A}}(\omega)$ is given by Eq.~(\ref{A_Bel_Th_Four_2}).

However, in the open-loop configuration ($\gamma = 0$), the noisy output power can be analytically determined as
\begin{eqnarray}
P_{0}=\frac{4 V_{\pi}^2 }{\pi R} \, \Delta F \, D_a \, ,
\label{noise_pow_gam0}
\end{eqnarray}
through the use the Fourier integral, or using fundamental results from stochastic calculus since Eq.~(\ref{A_Bel_Th}) degenerates to the well-known Orstein-Uhlenbeck equation.
Therefore, knowing the bandwidth  $\Delta F $ of the RF filter and the half-wave voltage $V_{\pi}$ of the MZ interferometer, an open-loop measurement of the output power can directly give an experimental a value for the white noise power density $D_a$ through Eq.~(\ref{noise_pow_gam0}).

\begin{figure}
\begin{center}
\includegraphics[width=8cm]{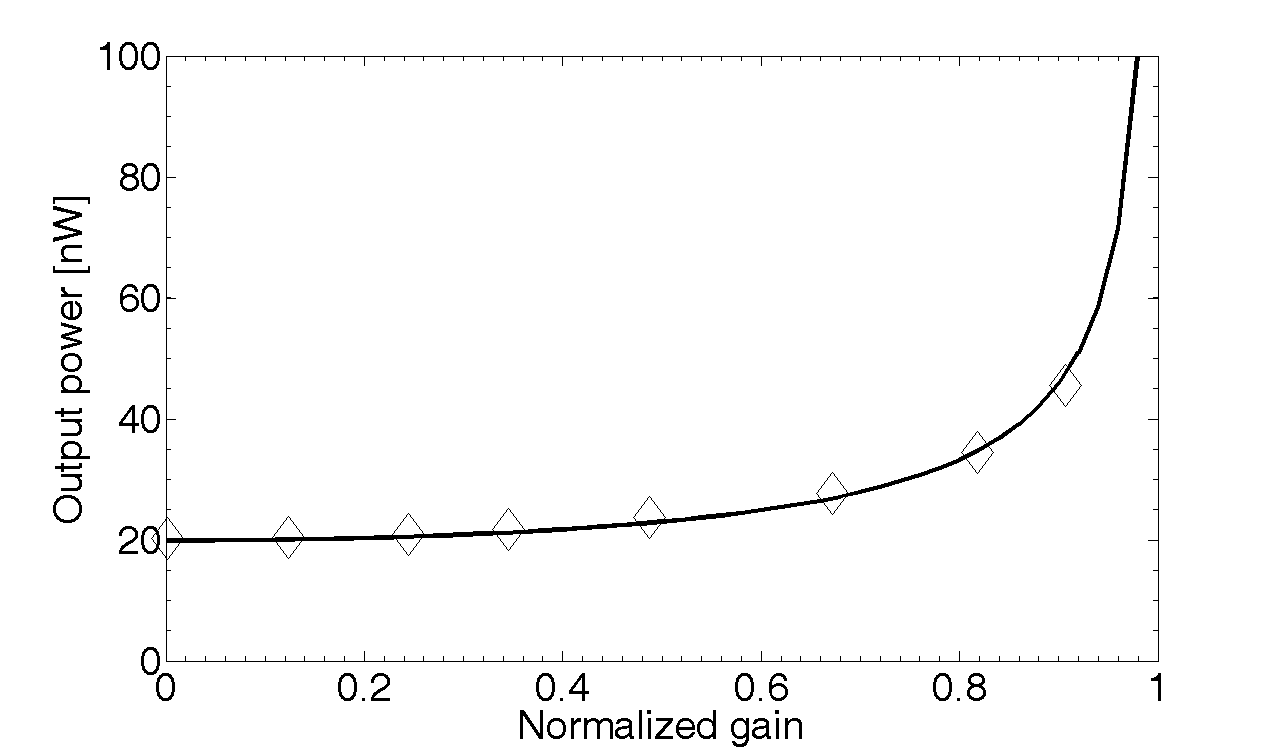}
\end{center}
\caption[Noise_Power_Profile]
{ \label{Noise_Power_Profile} Variation of the RF noise output power $P_{\gamma}$ as a function of the normalized gain, under threshold. The solid
line is the theoretical prediction of Eq. (\ref{noise_pow_true}) with $D_a= 9.8 \times 10^{-16}$ rad$^2$/Hz, and the symbols represent the experimentally measured data. The gain was varied through attenuation in the electric branch of the loop.}
\end{figure}

In our system, we have experimentally measured $P_{0}= 20$~nW (or $-47$~dBm), which corresponds to $D_a = 9.8 \times 10^{-16}$~rad$^2$/Hz.
This value for the power can also be obtained by other means [see Appendix A]. The curve displaying the power variation as a function of the normalized gain under threshold is shown in Fig.~\ref{Noise_Power_Profile}, and there is an excellent agreement between the experimental data and our analytical formula of Eq.~(\ref{noise_pow_true}). It may be interesting to note that the noise power apparently diverges at $\gamma =1$. In fact, one should not forget that this result is obtained using Eq.~(\ref{A_Bel_Th}), which is only valid for $|{\cal A}| \ll 1$. When $\gamma \rightarrow 1$, the amplitude of ${\cal A}$ increases and the higher order terms of the Bessel cardinal function are not negligible anymore, so that the Eqs.~(\ref{A_Bel_Th}) and (\ref{A_Bel_Th_Four_2}) are no more valid. Hence, divergence of the noise power is prevented by the nonlinear terms of Eq.~(\ref{eqt_A_stoch}) which become predominant in a very narrow range just below the threshold. A noteworthy study on this topic of noisy oscillators near threshold is ref. \cite{LaxVI}.

\begin{figure}
\begin{center}
\includegraphics[width=8cm]{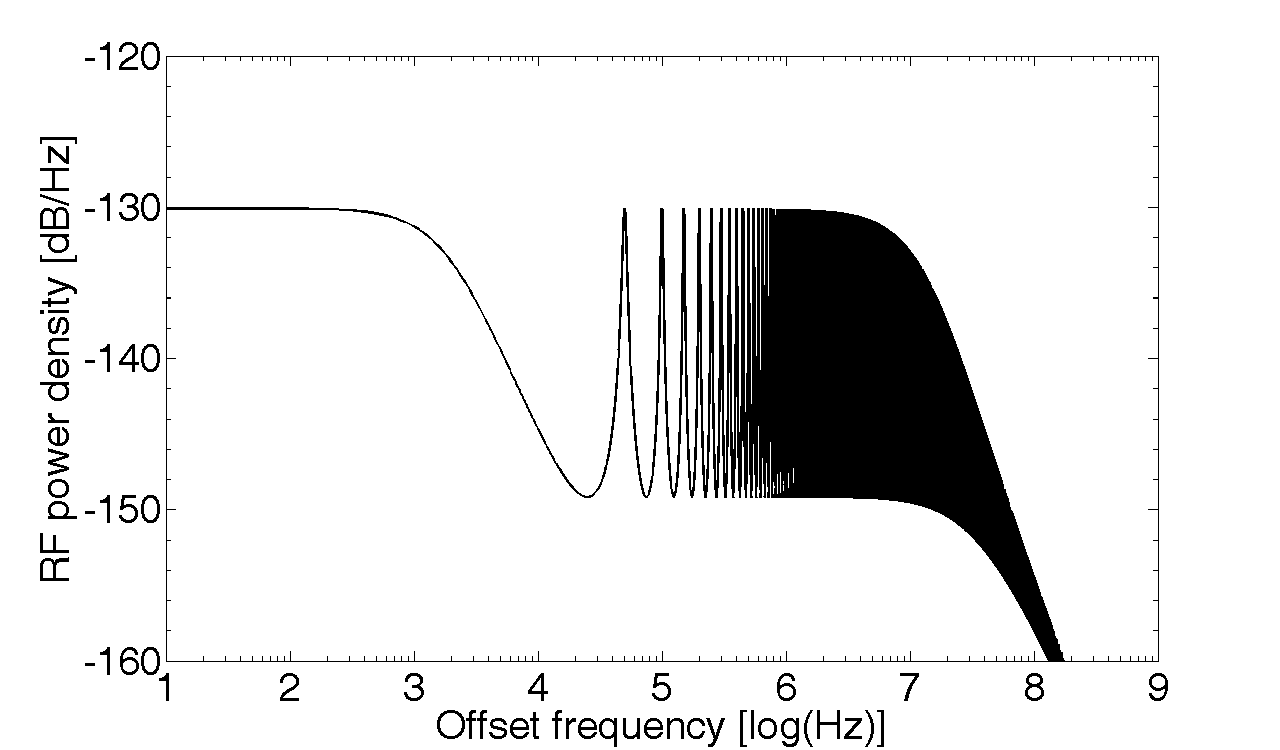}
\end{center}
\caption[Spec_BT_num]
{ \label{Spec_BT_num} Theoretical power density spectrum  $|\tilde{{\cal A}}(f)|^2$ of the microwave noise signal below threshold, with
$D_a= 9.8 \times 10^{-16} $ rad$^2$/Hz and $\gamma = 0.8$, using Eq. (\ref{A_Bel_Th_Four_2}). The semi-logarithmic scale is adopted because it enables to see at the same time the fine structure of regularly spaced ring-cavity peaks, and the global variation shaped by the RF filter bandwidth. This spectrum is divided into two areas: a quasi-flat area within bandwidth, and a $-20$ dB/dec decrease outside the bandwidth. }
\end{figure}

\begin{figure}[t]
\begin{center}
\includegraphics[width=8cm]{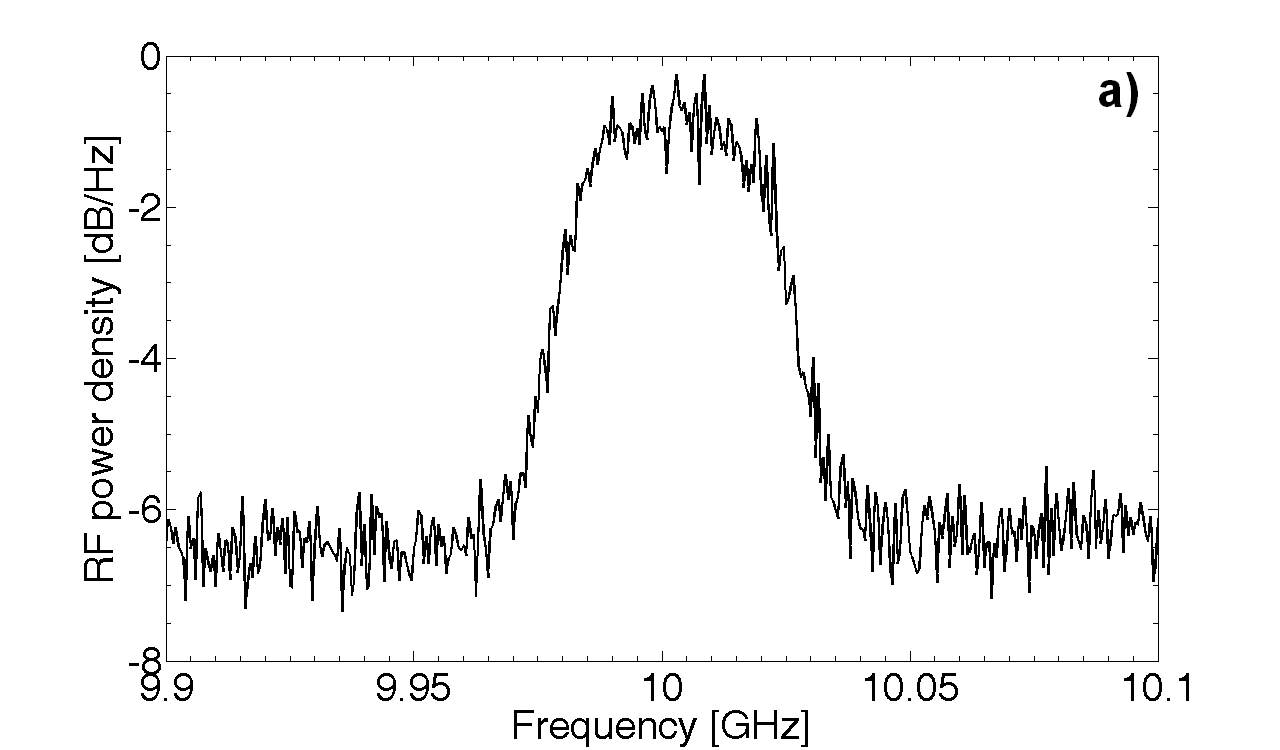}
\includegraphics[width=8cm]{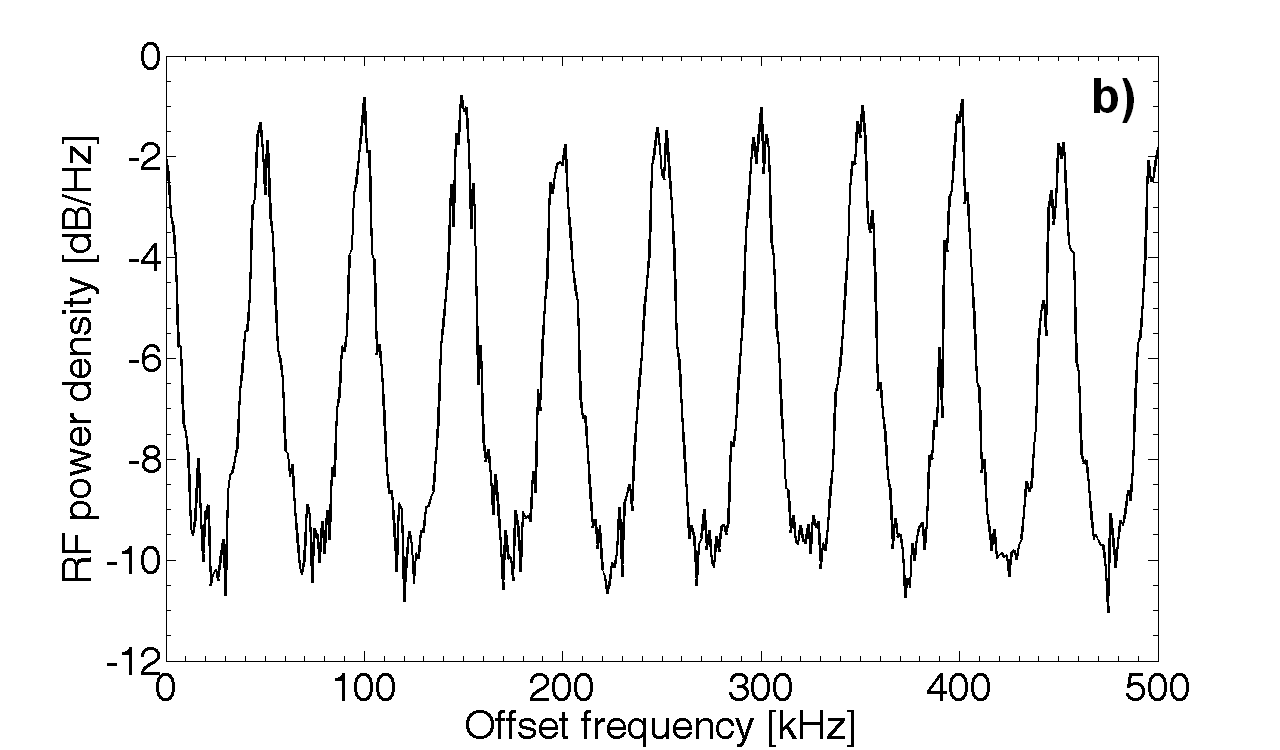}
\end{center}
\caption[Spec_BT_exp]
{ \label{Spec_BT_exp} An experimental power density spectra of the microwave noise below threshold. The spectrum has been scaled to its maximum (reference at $\sim 0$~dB). (a) Spectrum in a $200$~MHz window, showing how the noisy power density is profiled by the RF filter. (b) Zoom-in with near the $10$~GHz central frequency in a $500$~kHz window, showing the noisy ring-cavity peaks.}
\end{figure}

It is also noteworthy that for $\gamma = 0$, the noise spectrum follows the spectral shape of the RF filter. However, when $\gamma$ is increased (still below threshold), a first qualitative difference emerges, since the spectrum still follows the spectral shape of the filter, but its fine structure is composed by a collection of peaks which are the signature of microwave ring-cavity modes, as it can be seen in Figs.~\ref{Spec_BT_num} and \ref{Spec_BT_exp}.

\section{Phase noise spectrum above threshold (\boldmath$\gamma > 1$)}
\label{aboveth}

Above threshold, the amplitude of the microwave obeys the nonlinear algebraic equation ${\rm Jc_1}[2 |{\cal A}_0|]=1/(2 \gamma)$.
Linearizing Eq.~(\ref{eqt_A_stoch}) around this solution yields the following equation
\begin{eqnarray}
\dot{{\cal A}} = - \mu  e^{i\vartheta } {\cal A}
+ \mu e^{i\vartheta } \left[ 1+ \eta_m(t) \right] \, {\cal A}_T + \mu e^{i\vartheta } \zeta_a(t)\, .
\label{A_Above_Th}
\end{eqnarray}
We should now look for an equation for the phase $\psi$ in order to find its power density spectrum $|\Psi (\omega)|^2$.

\begin{figure}[t]
\begin{center}
\includegraphics[width=8cm]{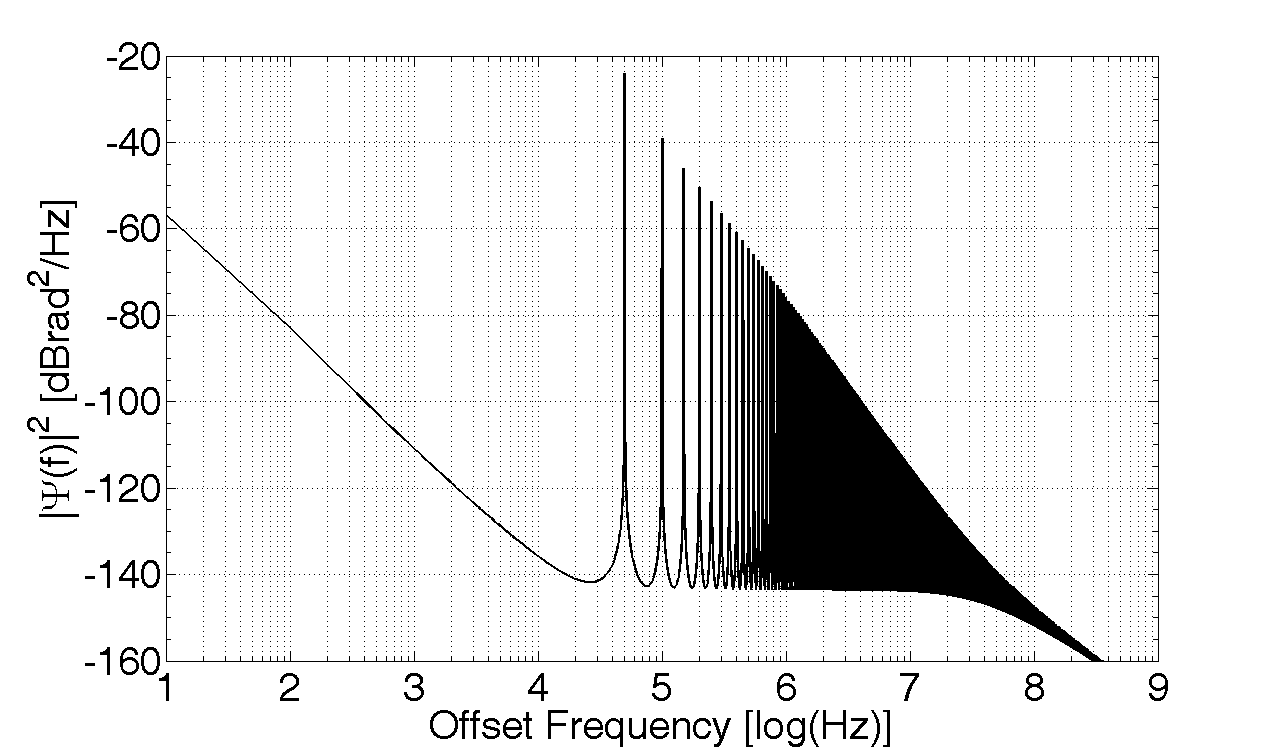}
\end{center}
\caption[Spec_AT]
{ \label{Spec_AT} A theoretical phase noise spectrum above threshold in a $500$~MHz window, with $D_a=9.8 \times 10^{-16}$~rad$^2$/Hz,
$D_m=5 \times 10^{-11}$~rad$^2$/Hz, $\Omega_{H}= 100$~kHz and $\Omega_{L}= 1$~Hz. The dimensionless amplitude of the microwave oscillation is
$|{\cal A}_0|=0.41$, corresponding to a power of $10.5$~dBm.}
\end{figure}

Using the It\^{o} rules of stochastic calculus [see Appendix B], we derive the following time-domain equation for the phase dynamics
\begin{eqnarray}
\dot{{\psi}} = - \mu (\psi -\psi_T) + \frac{\mu}{2Q} \, \eta_m(t) + \frac{\mu}{|{\cal A}_0|} \xi_{a,\psi}(t)\, ,
\label{eq_psi_Above_Th}
\end{eqnarray}
where $\xi_{a,\psi} (t)$ is a {\em real} Gaussian white noise of correlation $\left< \xi_{a,\psi}(t)\xi_{a,\psi}(t')\right> = 2 D_a \delta(t-t')$
(same variance as $\xi_{a} (t)$).
We can use Eq.~(\ref{eq_psi_Above_Th}) to obtain the Fourier spectrum $\Psi (\omega )$ of the phase $\psi (t)$, and then its power density spectrum following
\begin{eqnarray}
|\Psi (\omega)|^2 =  \left| \mu \, \frac{ (2Q)^{-1} \, \tilde{\eta}_m(\omega) + |{\cal A}_0|^{-1}\, \tilde{\xi}_{a,\psi}(\omega) }
                         { i \omega + \mu [1- e^{- i \omega T}]} \right|^2 \, .
\label{eq_psi_Ab_Th_Four}
\end{eqnarray}
Note that here, the influence of gain on phase noise spectrum is not explicit anymore: it is implicitly contained in $|{\cal A}_0|$.
Figure \ref{Spec_AT} displays the phase noise spectrum explicitly expressed by Eq. (\ref{eq_psi_Ab_Th_Four}), and we can now analyze how the spectrum behaves according to the various frequency ranges.

\subsection{Phase noise close to the carrier (\boldmath$\omega < \Omega_{H}$)}

Here, we consider the spectrum for frequencies which are relatively close to the carrier (with $\omega > \Omega_{L}$, however). Qualitatively, this corresponds to the frequencies that are much smaller than the high corner value $\Omega_{H}$ of the multiplicative flicker noise.
In this region, flicker noise is stronger than white noise, so that
$|\tilde{\eta}_m(\omega)|/2Q \gg |\tilde{\xi}_{a,\psi}(\omega)|$. On the other hand, we can also consider that $1-e^{- i \omega T} \simeq i\omega T$. Therefore, taking into account the fact that $\mu T \gg 1$, Eq.~(\ref{eq_psi_Ab_Th_Four}) can be simplified into
\begin{eqnarray}
|\Psi_{close} (\omega)|^2  \sim   \frac{|\tilde{\eta}_m(\omega)|^2}{4Q^2T^2} \, \frac{1}{\omega^2}
                   \simeq \frac{\Omega_{H}D_m}{2 Q^2 T^2} \, \frac{1}{\omega^3} \, .
\label{eq_psi_flicker}
\end{eqnarray}
Some remarks can be made at this stage. First, The Lesson effect is here very explicit: the phase noise spectrum decreases as $f^{-3}$ due to the $f^{-1}$ flicker noise \cite{rubiola}.
Secondly, The phase noise is inversely proportional to $Q^2$. Thirdly, the phase noise is practically independent of the microwave amplitude $|{\cal A}_0|$, as long as multiplicative noise is stronger than additive noise near the carrier.
Finally, the phase noise decreases as $T^{-2}$, therefore justifying the need for very long delay-lines to reduce phase noise close to the carrier.
This dependence was also recovered analytically by Yao and Maleki, using another theoretical approach \cite{YaoMalekiJOSAB}.
Hence, in first approximation there are three ways to reduce phase noise close to the carrier: reduce the power $D_m$ of the flicker noise,
increase the delay $T$ or increase the $Q$ factor of the RF filter.

\subsection{Phase noise in the spurious peaks range (\boldmath$\Omega_{H} < \omega \ll \mu$)}

The frequencies of concern are here those which are well within the bandwidth, but not too close from the carrier. This range typically lies between
$50$~kHz and few MHz, and contains the parasite ring-cavity peaks. It is also an area where the multiplicative and additive noises are both white
[in the sense that in that range, $|\tilde{\eta}_m(\omega)|^2$ and $|\tilde{\xi}_{a,\psi}(\omega)|^2$ are both constant].

The local minima of the spectrum in that area are obtained for $e^{-i \omega T}=-1$, so that the floor of the phase noise after the flicker decrease is
\begin{eqnarray}
|\Psi_{floor}|^2  \simeq \frac{1}{4}\,\left[ \frac{\sqrt{2 D_m}}{2Q} + \frac{\sqrt{2 D_a}}{|{\cal A}_0|}\right]^2
                  \sim  \frac{1}{2} \frac{D_a}{|{\cal A}_0|^2} \,
\label{eq_psi_floor}
\end{eqnarray}
when $Q$ is sufficiently high. This level is $6$ dB below the additive white noise power density scaled to the power of the microwave. The recipe for a low phase noise floor is then quite simple, and also quite conventional: low additive noise $D_a$,
and high power $|{\cal A}_0|^2$ for the microwave signal. The bandwidth does not play any role in this case, as long as the multiplicative noise is not too strong.

\begin{figure}[t]
\begin{center}
\includegraphics[width=8cm]{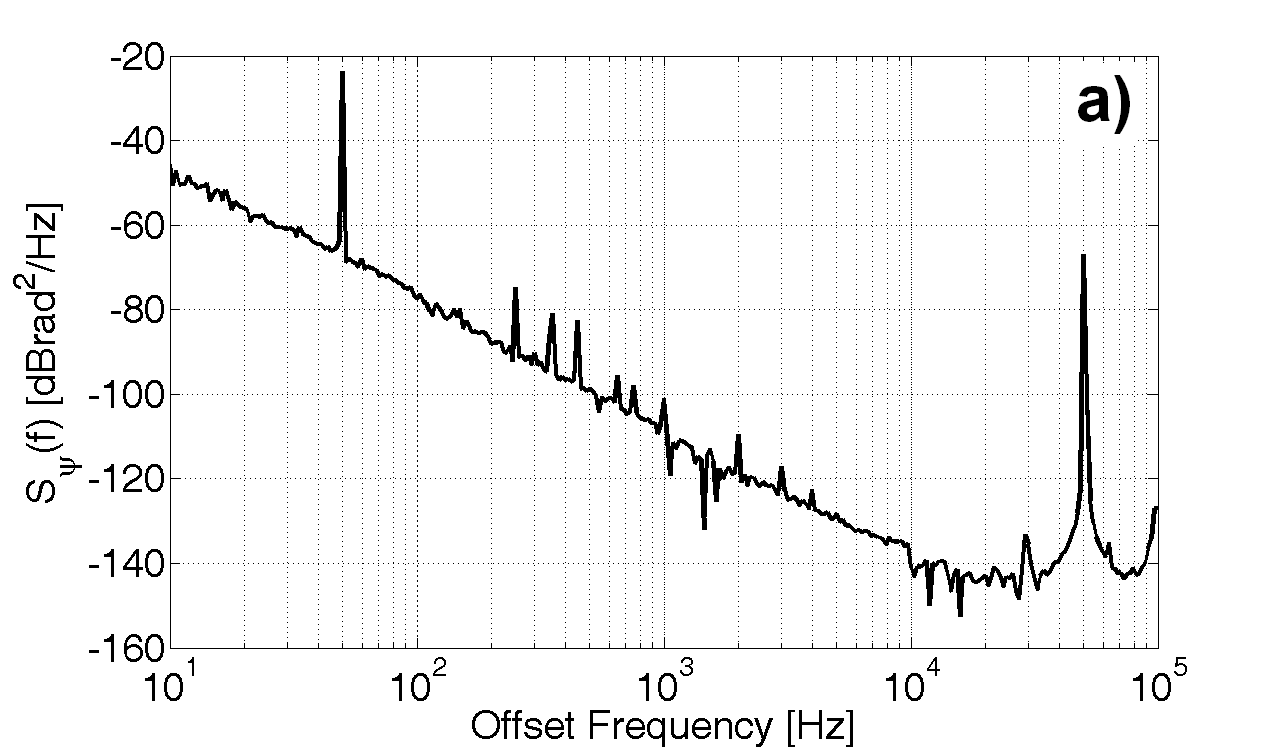}
\includegraphics[width=8cm]{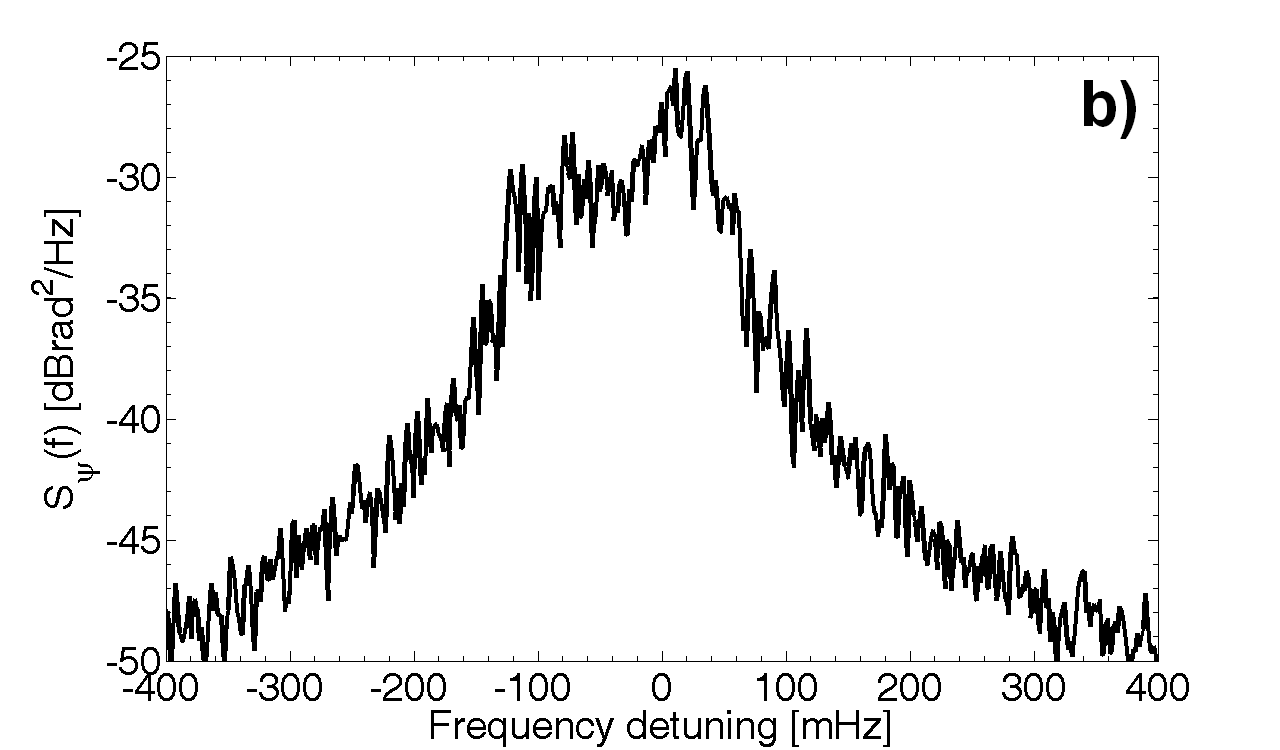}
\end{center}
\caption[Spec_AT_exp]
{ \label{Spec_AT_exp} $a)$ Experimental phase noise spectrum in a $100$~kHz window, showing a noise floor around $-145$~dBrad$^2$/Hz for a microwave power of $P=10.5$~dBm. $b)$ Enlargement of the spectrum around the first spurious peak at the frequency $f_1= 50594.35$~kHz. The maximum of this spurious peak is at $-25.5$~dBrad$^2$/Hz [height of the peak: $-119.5$~dB], and its $-3$~dB bandwidth is around $40$~mHz. All these experimental data are in excellent agreement with the theory. Note that the height of the peak in Fig. $a)$ is not indicative because of insufficient resolution. Also note that the peak at $50$ Hz is a parasite peak originating from the electric mains supply.}
\end{figure}

The spurious peaks are localized around integer multiples of the round-trip frequency $\Omega_T/2\pi=1/T=50$~kHz.
More precisely, a fourth order Taylor expansion of the denominator of Eq.~(\ref{eq_psi_Ab_Th_Four}) shows that around these resonance frequencies, the phase noise can be expressed as
\begin{eqnarray}
&& |\Psi (n\Omega_T + \delta \omega)|^2 =  \label{eq_psi_expand}  \\
&&  \frac{\mu^2 \left| (2Q)^{-1} \, \tilde{\eta}_m(\omega) + |{\cal A}_0|^{-1}\, \tilde{\xi}_{a,\psi}(\omega) \right|^2}
{ (n\Omega_T + \mu T \delta \omega)^2 - \frac{1}{3}n \Omega_T \mu T^3 \delta \omega^3 - \frac{1}{12}\mu^2 T^4 \delta \omega^4}  \nonumber \, .
\end{eqnarray}
By finding the minima of this Taylor-expanded denominator, it can be shown that the spurious peaks are in fact frequency-shifted according to
\begin{eqnarray}
f_{n}=  \frac{n}{T}  - \frac{n}{\pi \Delta F \, T^2} = n \times 50 \, {\rm{kHz}}- n \times 16 \, {\rm{Hz}} \, .
\label{eq_freq_spur}
\end{eqnarray}
Then, their height relatively to the phase noise floor can also be calculated as
\begin{eqnarray}
\Delta |\Psi_{n}|_{{\rm dB}}^2= 10 \log \left[ \frac{\Delta F \, T}{n}  \right]^4 = 120 \, {\rm{dB}} - 40 \log n \, .
\label{eq_psi_peaks}
\end{eqnarray}
It appears that the level of the spurious peaks increases with the RF bandwidth and with the delay: therefore, a large delay may lead to a lower phase noise near the carrier [see Eq.~(\ref{eq_psi_flicker})], but it also leads to a higher level for the spurious peaks, so that an optimal trade-off has to be found. It is also noteworthy that this level is {\em independent} of the power densities $D_a$ or $D_m$. In our case, the height of the first spurious peak relatively to the floor is theoretically equal to $120$~dB, in excellent agreement with the experimental results of Fig. \ref{Spec_AT_exp}, where a height of $119.5$~dB has been measured. It can also be shown from Eq.~(\ref{eq_psi_expand}) that the -$3$~dB bandwidth of the spurious peaks is
 \begin{eqnarray}
\Delta f_n = \frac{2}{\pi} \frac{n^2 }{\Delta F^2 T^3} = n^2 \times 32 \, \rm{mHz} \, ,
\label{eq_peaks_BW}
\end{eqnarray}
an extremely small value which is experimentally confirmed with the results of Fig. \ref{Spec_AT_exp}b. In comparison, these spurious peaks typically have a linewidth higher than $1$~kHz below threshold (see Fig. \ref{Spec_BT_num}), but their linewidths sharply narrow as $\gamma \rightarrow 1$.

\subsection{Phase noise outside the bandwidth (\boldmath$\omega > \mu$)}

Here, the term $\mu [1- e^{- i \omega T}]$ progressively becomes negligible as $\omega$ is increasing, so that
the ring-cavity peaks excited by white noise become strongly damped (for being outside the RF bandwidth). In this case, the phase noise decays as
\begin{eqnarray}
|\Psi_{out}(\omega)|^2 \simeq \frac{ 2 D_a \mu^2}{|{\cal A}_0|^2} \, \frac{1}{\omega^2}
                       \simeq \frac{\Delta F^2 D_a}{2 |{\cal A}_0|^2} \, \frac{1}{\omega^2} \, .
\label{eq_psi_out}
\end{eqnarray}
However, the phase noise does not decrease monotonically as $f^{-2}$ up to infinity: in fact, for $\omega \gg \mu$, there is a second phase noise floor induced by the coupling between phase fluctuations and amplitude fluctuations (second-order effect, see ref. \cite{LaxV}).

\section{Conclusion}
\label{conclusion}

This article has presented a theoretical study of phase noise in OEOs. Our approach has consisted in a Langevin formalism, that is, in adding noise sources to a core deterministic model for the microwave dynamics. We have found a excellent agreement between the main predictions of the model and the experimental results. There is also an good agreement between this theory and the results that are known from the literature, or from our earlier works.

The main advantage of this approach is that it enables within the same framework to understand the behavior of the system under and above threshold,
as the same model continuously accounts for all the observed features independently of the value of the gain.
However, we have not taken into account in this first model the noise generated by the filter (noisy $\mu$ and $\vartheta$), and the delay time (noisy $T$). Fluctuations associated to these parameters may induce interesting stochastic features, that will be adressed in future work.
Another line of investigation is to achieve a better spectral and statistical fitting of the multiplicative noise $\eta_m (t)$, which is an essential variable for the determination of phase noise spectra. Future work will also emphasize on phase noise reduction methods, such as optical filtering, multiple-loop architectures, or quadratic crossed nonlinearities \cite{Pomeau}.

\appendix

\section{Determination of the output noise power for \boldmath$\gamma =0$}

In the open-loop configuration, the total output power can also be obtained using some quantum electronics formulas.

Effectively, the output power can be explicitly expressed as
\begin{eqnarray}
P_0= \frac{1}{2}\,[FkT_0+2eI_{ph} R_{eq}]\,G \Delta F \, ,
\label{app_1}
\end{eqnarray}
where $G$ is the total gain of our two cascaded amplifiers ($22.5$ and $22.3$~dB at $10$~GHz), $F=6$~dB is the noise figure of the first amplifier,
$T_0=295$~K is the room temperature, $k$ is the Boltzmann constant, $e$ is the electron charge, $I_{ph}=1.2$~mA is the photodiode current,
$R_{eq}=25$~$\Omega$ is the equivalent load impedance for the photodiode, and $\Delta F= 50$~MHz is the bandwidth of the RF filter. The formula gives
$P_0= 19.6$~nW, while we have measured $20.0$~nW. The combination of Eqs. (\ref{noise_pow_gam0}) and (\ref{app_1}) also gives a method to determine $D_a$ directly from the specifications of the various optoelectronic components used in the oscillation loop.

\section{Derivation of the stochastic phase equation}

We use It\^o chain rules to derive the stochastic differential equation for the phase. We first rewrite Eq. (\ref{A_Above_Th}) under the differential form
\begin{eqnarray}
d {\cal A}= - \mu  e^{i\vartheta } {\cal A} dt +   \mu e^{i\vartheta } [1+ \eta_m (t) ] {\cal A}_T dt + \mu e^{i\vartheta }
              \, d{\cal W}_a \, ,
\label{app_9}
\end{eqnarray}
where $d{\cal W}_a(t)= dW_{a,r}(t)+i dW_{a,i}(t)$ is a differential Wiener process. Note that $\left< dW_{a,r} \right>=\left< dW_{a,i} \right>=0$ and
$\left< (dW_{a,r})^2 \right>=\left< (dW_{a,i})^2 \right> =2D_a dt$ . The fact that
$dW \sim {\cal O}(\sqrt{dt})$ explains why the
differential terms of second order should be taken into account in stochastic calculus, so that usual differentiation and chain rules do not generally apply. When considering the second order one may consider $(dW_{a})^2 \approx  \left< (dW_{a})^2\right> =2D_a dt$, and discard higher order terms since $\left<(dW_a)^{k+2}\right> \approx {\cal O}[(dt)^k] \ll dt$ for $k>0$.

We set ${\cal A}(t)= \sqrt{{\cal P}(t)} \, e^{i \psi(t)}= e^{\rho(t) + i \psi(t)}$, where $ \rho = \frac{1}{2} \ln {{\cal P}}$ is an auxiliary variable.
At order $dt$, we have
\begin{eqnarray}
d \rho +i d\psi &=& d \ln {{\cal A}} = \frac{d{\cal A}}{{\cal A}} -\frac{1}{2} \, \left[ \frac{d{\cal A}}{{\cal A}} \right]^2   \nonumber \\
                &=& - \mu  e^{i\vartheta } \, dt + \mu e^{i\vartheta } [1+ \eta_m (t) ]\, {\frac{{\cal A}_T}{{\cal A}}} \,dt \nonumber \\
				&&		+ \frac{\mu e^{i\vartheta }}{{{\cal A}}}   \, d{\cal W}_a   \, ,
\label{app_4}
\end{eqnarray}
where we have considered $(d{\cal W}_a)^2 \approx \left<(d{\cal W}_a)^2\right> =0$. Assuming second order fluctuations for the amplitude
(that is, $|{\cal A}_T| \simeq |{\cal A}| $), we are led to
\begin{eqnarray}
d\psi &=& - \mu \sin \vartheta  \, dt + \mu [1+ \eta_m (t) ]\,\sin[\vartheta + \psi_T -\psi] \,dt \nonumber \\
	  &&		+ \frac{\mu}{|{\cal A}_0|}   \, dW_{a,\psi}   \, ,
\label{app_11}
\end{eqnarray}
where $d W_{a,\psi} = dW_{a,r} \sin \vartheta  + dW_{a,i} \cos \vartheta $ is a real Gaussian white noise with zero mean and variance
$\left<(d W_{a,\psi})^2\right>= 2D_a dt$. Since $|\psi - \psi_T| \ll \vartheta \ll 1$, we have
$\sin[\vartheta + \psi_T -\psi] \simeq \vartheta - (\psi -\psi_T)$ so that finally,
\begin{eqnarray}
\dot{{\psi}} = - \mu (\psi -\psi_T) + \frac{\mu}{2Q} \, \eta_m(t) + \frac{\mu}{|{\cal A}_0|}  \, \xi_{a,\psi}(t)\, .
\label{app_12}
\end{eqnarray}
This result is also the one we may have recovered through the usual rules of differential calculus (however, note that it is not so for the equation ruling the power variable ${\cal P}$). Also note that this equation is valid only as long as the approximation of neglecting $ \ddot{{\cal B}}$ in
Eq.~(\ref{eq_calB_stoch_1}) is valid.

\section{An alternative paradigm for phase noise analysis}

It is possible to gain a different physical insight into the phase noise problem in OEOs, using an alternative methodology related to the conventional theory of feedback oscillators. We hereafter briefly sketch the main lines of this heuristical approach.

\begin{figure}
\begin{center}
\includegraphics[width=8cm]{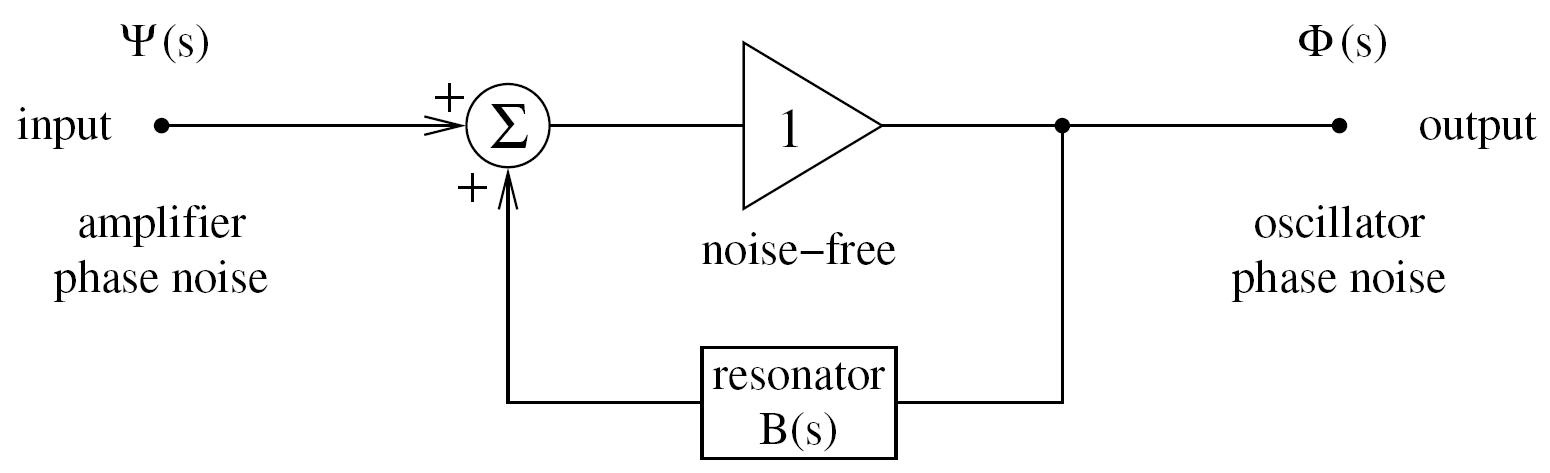}
\end{center}
\caption[fig:le-calc-dly-hphase]
{ \label{Scheme_Barkhausen} Oscillator phase noise transfer function (with $s \equiv jf$).}
\end{figure}

The oscillator consists of an amplifier of gain $A$ (constant) and of a feedback path of transfer function $\beta(jf)$ in closed loop. The function $\beta(jf)$ selects the oscillation frequency, while the gain $A$ compensates for the feedback loss. This general model is independent of the nature of the amplifier and of the frequency selector. We assume that the Barkhausen condition $|A\beta(jf)|=1$ for stationary oscillation is verified at the carrier frequency $f_0$ by through a gain-control mechanism.  Under this hypothesis, the phase noise is modeled by the scheme shown in Fig.~\ref{Scheme_Barkhausen}, in which all signals are \emph{the phases of the oscillator loop} \cite{rubiola}. The main reason for describing the oscillator in this way is that we get rid of the non-linearity, pushing it in the loop-gain stabilization. The ideal amplifier ''repeats" the phase of the input, for it has a gain of $1$ (exact) in the phase-noise model. The real amplifier introduces the random phase $\psi(t) \leftrightarrow \Psi(jf)$ in the loop. In this representation, the phase noise is always additive noise, regardless of the physical mechanism involved. This eliminates the mathematical difficulty inherent in the parametric nature of flicker noise and of the noise originated from the environment fluctuations.

The feedback path is described by the transfer function $\mathrm{B}(jf)$ of the phase perturbation.  In the case of the delay-line oscillator, the feedback path is a delay line of delay $T$ followed by a selector filter. The latter is necessary, otherwise the oscillator would oscillate at any frequency multiple of $1/T$, with no preference. Implementing the selector as a bandpass filter (a resonator) of group delay $T_g$, the phase-perturbation response of the feedback path is
\begin{eqnarray}
\mathrm{B}(jf)=\frac{\exp(-j 2 \pi f T)}{1+j 2 \pi f T_g} \, .
\end{eqnarray}
We assume that all the phase perturbations in the loop are collected in the random function $\psi(t) \leftrightarrow \Psi(jf)$, regardless of the physical origin (amplifier, photodetector, optical fiber, etc.).  Denoting with $\varphi(t) \leftrightarrow \Phi(jf)$ the oscillator output phase, the oscillator is described by the phase-perturbation transfer function $\mathrm{H}(jf)= \Phi(jf)/\Psi(jf)$. By inspection on Fig.~\ref{Scheme_Barkhausen}, and using the basic equations of feedback, the oscillator transfer function reads
\begin{eqnarray}
\mathrm{H}(jf)&=& \frac{1}{1-\mathrm{B}(jf)}  \, ,
\label{app_c_1}
\end{eqnarray}
and the oscillator phase noise spectrum would be given by $S_\varphi(f)=|\mathrm{H}(jf)|^2S_\psi(f)$.

\end{document}